\documentclass{article}
\usepackage{spconf,amsmath,graphicx,hyperref}
\usepackage{amssymb}
\usepackage{amsfonts}
\usepackage{booktabs}
\usepackage{balance}
\usepackage{cite}

\newcommand{\HT}{\mathsf{H}}

\title{Generating Training Targets for Real-World Speech Enhancement \\via Close-to-Distant Microphone Projection\vspace{-1mm}}
\name{Tomohiro Nakatani, Rintaro Ikeshita, Naoyuki Kamo, Marc Delcroix, Shoko Araki\vspace{-1mm}}
\address{NTT, Inc., Japan\vspace{-1mm}}
\usepackage[absolute,showboxes]{textpos}
\TPMargin{5pt}
\newcommand{\copyrightstatement}{
\begin{textblock}{0.8}(0.1,0.01)
\noindent
\footnotesize
\copyright 2026 IEEE.  Personal use of this material is permitted.  Permission from IEEE must be obtained for all other uses, in any current or future media, including reprinting/republishing this material for advertising or promotional purposes, creating new collective works, for resale or redistribution to servers or lists, or reuse of any copyrighted component of this work in other works.
\end{textblock}
}
\begin{document}
\ninept
\copyrightstatement
\maketitle
\begin{abstract}
Training neural networks (NNs) for speech enhancement (SE) in distant speech-capturing scenarios requires paired distorted and clean reference speech signals. While such data are often generated through simulation, the mismatch between simulated and real recordings significantly limits SE accuracy. To address this issue, we propose Close-to-Distant microphone Projection (C2D projection), a method that generates paired data from real recordings captured by close and distant microphones. C2D projection estimates an optimal projection matrix that transforms close-microphone inputs into clean reference signals aligned with distant-microphone recordings, while simultaneously performing denoising. We show this projection can be effectively realized using a variant of the Parametric Multichannel Wiener Filter (PMWF). Experimental results demonstrate that an NN trained with C2D-projected data outperforms the state-of-the-art Guided Source Separation (GSS) on the challenging CHiME6 dinner party ASR task under oracle diarization, when using the enhanced output from GSS as an auxiliary input to the NN.
\end{abstract}
\section{Introduction}
\label{sec:intro}
Speech captured with Close Microphones (CMs), positioned near the speaker’s mouth, generally provides high-quality recordings even in real-world environments. However, their intrusive nature can limit user comfort and natural interaction. In contrast, Distant Microphones (DMs), placed farther away, offer greater freedom and flexibility but often suffer from severe degradation due to reverberation, background noise, and interference from other speakers. Such degradation reduces speech intelligibility for human listeners and significantly impairs the performance of speech-based applications, including Automatic Speech Recognition (ASR).

Speech Enhancement (SE) aims to recover high-quality speech from such degraded DM signals. Signal processing-based and Neural Network (NN)-based approaches have been extensively studied \cite{reinhold2025spm}. Signal processing-based SE (e.g., \cite{SDW-MWF,Souden2007taslp,ono2010auxica,nakatani2021blind}) relies on assumptions about room acoustics and signal characteristics, typically requiring no prior training. This enables them to adapt well to diverse, unknown test conditions. In contrast, NN-based SE \cite{wang2022tfgridnet_2022,sepformer,gu2022towards} learns a mapping from distorted to clean speech and can achieve superior performance when training and test conditions are well matched.

Despite the success of NN-based SE in delivering high-quality results, its application to real-world recording scenarios, such as the CHiME6 dinner party conversations \cite{CHiME6_2020}, remains a major challenge. A key obstacle is the requirement for paired distorted and clean reference signals for NN training, which are difficult to obtain in realistic conditions. Simulation-based approaches, such as the image method \cite{imagemethod}, are widely used to generate paired data. While effective in relatively simple environments \cite{wang2022tfgridnet_2022}, these techniques have yet to prove successful in complex real-world scenarios. For example, in CHiME6, top-scoring systems rely on a signal processing-based Guided Source Separation (GSS) approach \cite{boeddeker2018chime5}, which combines Weighted Prediction Error (WPE) dereverberation \cite{WPE,MIMO-WPE} and Parametric Multichannel Wiener Filter (PMWF) \cite{Souden2007taslp}. Replacing GSS with more advanced NNs has had little success to date \cite{cornell2025recent}.

This paper proposes a novel approach, Close-to-Distant Microphone Projection (C2D projection), to address this challenge. The method assumes that both DM and CM signals are available during NN training, with CM signals being much less distorted than DM signals (noting that only DM signals are accessible at test time). This assumption is realistic for real-world datasets; for example, worn microphone recordings are provided as part of the training data in CHiME6. However, such CM signals cannot be directly used as training targets, since they are not phase- or power-aligned with DM signals and often contain interfering signals. The goal of C2D projection is thus to estimate a projection matrix that optimally aligns CM signals with DM signals while simultaneously denoising them to produce suitable training targets. We show that such a matrix can be derived as a variant of PMWF. Additionally, we explain how to apply the C2D projection to NN training using the CHiME6 scenario as the testbed. We show that the GSS approach, with minor modifications, can provide all the required techniques for the application. 

Experiments show that an NN trained with C2D-projected data outperforms the GSS approach on the CHiME6 ASR task under oracle diarization, when using the GSS-based PMWF output as an auxiliary input. Moreover, using the CHiME8 task \cite{CHiME8_2024} with the estimated diarization labels, the proposed method still achieves improvements over the GSS approach across most ASR scenarios, despite substantial mismatches with the training conditions, including differences in diarization labels (oracle vs. estimated) and diverse recording environments.

\section{Related work}

To achieve real-world SE, researchers have explored NN training techniques that do not rely on target generation \cite{lucus2019icassp, Nana2019apsipa, WisdomTEWWH20, bando2025icassp, ZQ_SuperM2M, Lei-AA2025}. Spatial self-supervised learning \cite{bando2025icassp} enables unsupervised NN training using only observed signals and has shown marginal ASR improvements over the GSS approach on the multi-source CHiME-8 task. SuperM2M \cite{ZQ_SuperM2M} and FNSE-SAT \cite{Lei-AA2025} enable the direct use of CM signals as training targets. The former combines unsupervised learning on real data with supervised learning on simulated data, while the latter employs supervised adversarial training. Both techniques have demonstrated improved ASR performance on real-world single-source tasks.


In contrast to these techniques, our method generates training targets directly from real data. This allows us to use a simple supervised learning scheme for NN training and results in clear effectiveness on the challenging multi-source CHiME-6 task.

\section{Proposed method}
\label{sec:format}
After defining the problem to be solved, this section describes the proposed method, C2D projection. 

\subsection{Problem definition}
Suppose that the speech signals of $N$ speakers are captured by $M^c$ CMs positioned near the speakers, and $M^d$ DMs. Let $s_n(t,f)\in\mathbb{C}$ denote the $n$th speaker's clean speech signal in the short-time Fourier transform (STFT) domain, where $t$ and $f$ are the time and frequency indices, respectively. Let $\mathbf{x}^c(t,f)\in\mathbb{C}^{M^c}$ and $\mathbf{x}^d(t,f)\in\mathbb{C}^{M^d}$ denote vectors of signals captured by all CMs and DMs, respectively. Then, we model the signals by:
\begin{align}
\left[\begin{array}{c}\mathbf{x}_n^c(t,f)\\\mathbf{x}_n^d(t,f)\end{array}\right]
&=\left[\begin{array}{c}\mathbf{h}_n^c(f)\\\mathbf{h}_n^d(f)\end{array}\right]{s}_n(t,f),\label{eq:image}\\
\left[\begin{array}{c}\mathbf{x}^{c}(t,f)\\\mathbf{x}^{d}(t,f)\end{array}\right]
&=\sum_{n=1}^N\left[\begin{array}{c}\mathbf{x}_n^c(t,f)\\\mathbf{x}_n^d(t,f)\end{array}\right]
+\left[\begin{array}{c}\mathbf{v}^c(t,f)\\\mathbf{v}^d(t,f)\end{array}\right],
\label{eq:obs}
\end{align}
where $\mathbf{h}_n^c(f)\in\mathbb{C}^{M^c}$ and $\mathbf{h}_n^d(f)\in\mathbb{C}^{M^d}$ are vectors containing acoustic transfer functions (ATFs) from the $n$th speaker to the CMs and DMs, and $\mathbf{x}_n^c(t,f)$ and $\mathbf{x}_n^d(t,f)$ are source images of the $n$th speaker's speech at the respective microphones. We assume that the ATFs are time-invariant within each short time segment of interest. Then, the captured signals, $\mathbf{x}^{c}(t,f)$ and $\mathbf{x}^{d}(t,f)$, are modeled by the sum of all source images and the additive noise, $\mathbf{v}^c(t,f)$ and $\mathbf{v}^d(t,f)$, at respective microphones. 

This paper assumes that the goal of training target generation is to estimate the source images of individual speakers at DMs, $\mathbf{x}_n^d(t,f)$, which serve as training targets corresponding to the observed DM signal $\mathbf{x}^d(t,f)$.
We adopt this objective based on prior studies demonstrating that using such paired data is effective for achieving high-quality speech enhancement in simulation experiments \cite{wang2022tfgridnet_2022,nakicassp2025}.

According to this goal, we set the goal of C2D projection to estimate a projection matrix ${\bf{W}}_n(f)\in\mathbb{C}^{M^c\times M^d}$ that transforms the CM signal $\mathbf{x}^c$ to an estimate of the $n$th speaker's source image at the DMs, $\hat{\mathbf{x}}_n^d$:
\begin{align}
\hat{\mathbf{x}}_n^d(t,f)=\mathbf{W}_n(f)^{\HT}\mathbf{x}^c(t,f),
\label{eq:filtering}
\end{align}
where $(\cdot)^{\HT}$ denotes the Hermitian transpose. With the projection matrix for each utterance, we can prepare paired data composed of the speech reference $\hat{\mathbf{x}}_n^d(t,f)$ and the distorted observation $\mathbf{x}^d(t,f)$ at the DMs, and use them for the training of the SE model. 

When the CM signal in Eq.~(\ref{eq:obs}) does not contain additive noise and all ATFs from the sources to the microphones are given, the exact projection matrix can be obtained, provided $M^c\ge N$, by:
\begin{align}
    \mathbf{W}_n(f)^{\HT}=\mbox{diag}\{\mathbf{h}_n^d(f)\}\mathbf{H}^c(f)^+
    \label{eq:solution}
\end{align}
where $\mathbf{H}^c(f)=[\mathbf{h}_1^c(f),...,\mathbf{h}_N^c(f)]$, $\mathbf{H}^+$ is the Moore-Penrose pseudo-inverse of a matrix $\mathbf{H}$, and $\mbox{diag}\{\mathbf{h}\}$ is a diagonal matrix containing elements of $\mathbf{h}$ as its diagonal components.  The issue to be solved in this paper is thus to optimally estimate $\mathbf{W}_n(f)$ with no prior knowledge of the ATFs in a way robust against additive noise in the CM and DM signals. 

\subsection{C2D projection based on a variant of PMWF}
This paper adopts the following cost function as the estimation criterion of the projection matrix for the $n$th source:
\begin{align}
C(\mathbf{W}_n(f))=&E\{\|\mathbf{W}_n(f)^{\HT}\mathbf{x}_n^c(t,f)-\mathbf{x}_n^d(t,f)\|_2^2\}\nonumber\\&+\mu E\{\|\mathbf{W}_n(f)^{\HT}\mathbf{v}_n^c(t,f)\|_2^2\},\label{eq:cost}\\
\mbox{where}~~\mathbf{v}_n^c(t,f)=&\sum_{n'\neq n}\mathbf{x}_{n'}^c(t,f)+\mathbf{v}^c(t,f).\label{eq:vnc}
\end{align}
The first term on the right-hand side in Eq.~(\ref{eq:cost}) quantifies the projection error between the DM source images and the transformed source images for the $n$th speaker. The second term quantifies the power of the unwanted signal $\mathbf{v}_n^c$, composed of additive noise and voices from other speakers, remaining after projection. $\mu$ is the weight that balances the two terms. Minimizing the cost function yields a projection matrix that jointly reduces projection error and unwanted signal in a balanced manner.

The above cost function is equivalent to that of the Speech Distortion Weighted Multichannel Wiener Filter (SDW-MWF) \cite{SDW-MWF}, except that the observed and the target signals are in different domains, i.e., CM and DM domains, in our formulation. As with the SDW-MWF, a closed-form solution that minimizes the cost function can be derived:
\begin{align}
\mathbf{W}_n(f)&=\left(\mathbf{\Phi}_n^{cc}(f)+\mu\mathbf{\Phi}_{v_n}^{cc}(f)\right)^{-1}\mathbf{\Phi}_n^{cd}(f),\label{eq:sdwmwf}\\
\mathbf{\Phi}_n^{cc}(f)&=E\{\mathbf{x}_n^c(t,f)\mathbf{x}_n^c(t,f)^{\HT}\},\label{eq:ncov}\\
\mathbf{\Phi}_{v_n}^{cc}(f)&=E\{\mathbf{v}_n^c(t,f)\mathbf{v}_n^c(t,f)^{\HT}\},\label{eq:vncov}\\
\mathbf{\Phi}_n^{cd}(f)&=E\{\mathbf{x}_n^c(t,f)\mathbf{x}_n^d(t,f)^{\HT}\}\label{eq:crosscov},
\end{align}
where $\mathbf{\Phi}_n^{cc}(f)$ and $\mathbf{\Phi}_{v_n}^{cc}(f)$ are spatial covariance matrices of the $n$th source and the unwanted signal in the CM domain, and $\mathbf{\Phi}_n^{cd}(f)$ is the cross-covariance matrix of the $n$th source across CM and DM domains. Further assuming $\mathbf{\Phi}_n^{cc}$ and $\mathbf{\Phi}_n^{cd}$ are both rank-1, which can be supported by Eq.~(\ref{eq:image}), we obtain the following formula for the C2D projection matrix:
\begin{align}
\mathbf{W}_n(f)=\frac{\mathbf{\Phi}_{v_n}^{cc}(f)^{-1}\mathbf{\Phi}_n^{cd}(f)}{\mu+\mbox{tr}\{\mathbf{\Phi}_{v_n}^{cc}(f)^{-1}\mathbf{\Phi}_n^{cc}(f)\}},\label{eq:c2d}
\end{align}
where $\mbox{tr}\{\cdot\}$ denotes matrix trace. This is equivalent to PMWF \cite{Souden2007taslp} except for $\mathbf{\Phi}_n^{cd}$ in the numerator, which enables the C2D projection in our formulation. 

It is important to note that the tradeoff between projection error reduction and noise suppression can be controlled by adjusting the parameter $\mu$, in a manner analogous to PMWF. For instance, setting $\mu = 0$ enforces the prediction error to be exactly zero, but at the cost of reduced noise suppression capability. This setting corresponds to the Minimum-Variance Distortionless Response (MVDR) BF within the PMWF framework.

A remaining issue in applying the C2D projection is how to estimate $\mathbf{\Phi}_{v_n}^{cc}$, $\mathbf{\Phi}_n^{cc}$, and $\mathbf{\Phi}_n^{cd}$ in Eq.~(\ref{eq:c2d}).
Various techniques can be used depending on the recording scenarios, and a representative case based on the CHiME6 setup will be introduced in the next section. 

\section{Application to CHiME6 scenario}

This paper adopts the CHiME6 scenario as the testbed, as it closely matches the assumptions underlying the C2D projection. In this section, we first provide a brief overview of the CHiME6 scenario, then describe how the C2D projection can be applied to it.

\begin{figure}[t]
    \begin{center}
\includegraphics[width=0.75\linewidth]{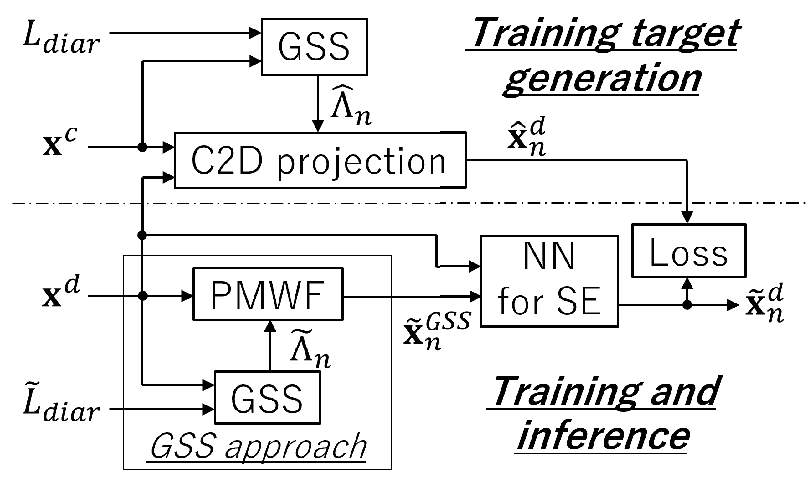}~\vspace{-3mm}\\
    \end{center}
\caption{Processing flow of C2D projection-based training target generation and NN training/inference for SE in the CHiME6 scenario.}\label{fig:flow}
\end{figure}

Figure~\ref{fig:flow} illustrates the processing pipeline for C2D projection-based training target generation, and the training and inference stages of the NN for SE.

\subsection{CHiME6 scenario}
In the CHiME6 scenario, dinner-party conversations by $N=4$ speakers were recorded using both worn microphones and microphone arrays. Each speaker wore a pair of microphones, yielding 
$M^c = 8$ CMs (4 speakers × 2 worn mics). Six microphone arrays were distributed throughout the environment, each with four microphones, yielding 
$M^d = 24$ DMs (6 arrays × 4 mics).

Manually generated oracle diarization labels $L_{diar}$ are provided, indicating the start and end times of each utterance for each speaker. The dataset comprises 20 conversation sessions, each approximately one hour long, split into 16 for training (train), 2 for development (dev), and 2 for evaluation (eval).

An effective SE baseline for CHiME6 is the GSS approach \cite{boeddeker2018chime5}. In this baseline, time–frequency (TF) masks $\tilde{\Lambda}_n(t,f)$, which indicate the dominant source at each TF point, are first estimated using GSS by guiding the complex Angular Central Gaussian Mixture Model (cACGMM) \cite{ito2016complex} with estimated diarization labels $\tilde{L}_{diar}$.  These masks are then used to compute the covariance matrices of the target and unwanted signals, and the enhanced target is finally obtained using the PMWF \cite{Souden2007taslp}.

The goal is to estimate who speaks when and what, using only the DM signals and models trained on the training data. The CM signals and oracle diarization labels are assumed to be available only during training.

\subsection{Estimation of C2D projection matrix}
For estimating the C2D projection matrix, all required techniques are already available within the GSS approach, with only minor modifications needed. Additionally, since the C2D projection is used to generate training targets, all data, including CM signals and oracle diarization labels, can be utilized for the estimation.

Unlike the original GSS approach, which estimates both the TF masks and the PMWF from the DM signals, our proposed method derives the TF masks from the CM signals and estimates the projection matrix using both CM and DM signals. The processing flow is summarized:
\begin{enumerate}
\item Extract each segment, which contains a target speaker’s utterance and unwanted signals, using the oracle diarization labels.
\item Apply GSS, using the labels as guidance, to the CM signals for each segment, and obtain TF masks $\hat{\Lambda}_n(t,f)$ for the target speaker's utterance.
\item Calculate the covariance matrices for the utterance:
\begin{align}
\mathbf{\Phi}_n^{cc}&=\frac{\sum_t \hat{\Lambda}_n(t,f)\mathbf{x}^c(t,f)\mathbf{x}^c(t,f)^{\HT}}{\sum_t \hat{\Lambda}_n(t,f)},\label{eq:ncovcal}\\
\mathbf{\Phi}_{v_n}^{cc}&=\frac{\sum_t (1-\hat{\Lambda}_n(t,f))\mathbf{x}^c(t,f)\mathbf{x}^c(t,f)^{\HT}}{\sum_t (1-\hat{\Lambda}_n(t,f))},\label{eq:vncovcal}\\
\mathbf{\Phi}_n^{cd}&=\frac{\sum_t \hat{\Lambda}_n(t,f)\mathbf{x}^c(t,f)\mathbf{x}^d(t,f)^{\HT}}{\sum_t \hat{\Lambda}_n(t,f)},\label{eq:crosscovcal}
\end{align}
\item Calculate and apply $\mathbf{W}_n(f)$ by Eqs.~(\ref{eq:c2d}) and (\ref{eq:filtering}) to obtain a training target $\hat{\mathbf{x}}_n^d(t,f)$ for the DM signal $\mathbf{x}^d(t,f)$.
\end{enumerate}

In the above procedure, the three types of covariance matrices are all derived from the observed CM and DM signals using TF masks in Eqs.~(\ref{eq:ncovcal})–(\ref{eq:crosscovcal}). Similar to the GSS approach, we here assume that unwanted signals present in the CM signals can be effectively suppressed by using the masks as the weighting function. Additionally, we assume that unwanted signals present only in the DM signals, $\mathbf{x}^d(t,f)$, are uncorrelated with the CM signals, $\mathbf{x}^c(t,f)$, and can therefore be disregarded in Eq.~(\ref{eq:crosscovcal}).

\subsection{Training of NN-based SE model for CHiME6}
In the CHiME6 scenario, each segment of the DM signals typically contains overlapping speech, necessitating a mechanism for the SE model to identify the target speaker. To address this, as shown in Fig.~\ref{fig:flow}, we utilize the GSS-based PMWF output $\tilde{\mathbf{x}}_n^{GSS}$ as auxiliary input to the SE model. The SE model is thus trained with both the raw DM signal ${\mathbf{x}}^d$ and the PMWF output $\tilde{\mathbf{x}}_n^{GSS}$ as observed inputs. The model's output $\tilde{\mathbf{x}}_n^d$ is then compared with the training target $\hat{\mathbf{x}}_n^d$ generated by the C2D projection, and the resulting loss is used to train the model.

To fully exploit the information in multichannel observations, we employed a Multi-Input Multi-Output (MIMO) SE configuration. Specifically, we selected $M(=4)$ channels from both the observed DM signals and the GSS output to serve as inputs to the SE model.  The corresponding channels from the generated training targets were used as the training reference. These channels were selected based on signal-to-noise ratios (SNRs) estimated from the DM signals using the method provided in the CHiME8 baseline system \cite{boeddeker2018chime5}. 

For the SE model, we adopted Deterministic Recursive Enhancement (DRE), a simplified variant of Probabilistic-Deterministic Recursive Enhancement (PDRE) \cite{nakicassp2025}. While PDRE estimates both the enhanced speech and its distribution, DRE removes the distribution estimation path from PDRE and estimates only the enhanced speech. We choose DRE for its simplicity and its ability to deliver comparable performance to PDRE. 
To perform MIMO SE, we adopted the multi-stream Noise Conditional Score Network (mNCSN++) \cite{nakiwaenc2024} as the enhancement network, which is recursively applied within DRE. 

\section{Experiments}

This section demonstrates the effectiveness of the proposed method through experiments on the CHiME8 dataset \cite{CHiME8_2024}. The NN for SE was trained solely on the CHiME6 train set using oracle diarization labels, and evaluated under various conditions. The evaluation covered all four different recording scenarios in the CHiME8 dataset, including CHiME6, Dipco, Mixer6, and Notsofar1, under estimated diarization, which contains substantial mismatches with the training condition.

\subsection{Analysis condition}\label{sec:acond}
For training the NN for SE, we generated data pairs for all sessions in the CHiME6 train set using the C2D projection with oracle diarization labels. Of these data pairs, 14 sessions were used for training, and the remaining two sessions were reserved for validation. DRE for the NN was implemented using publicly available code\footnote{https://github.com/sp-uhh/sgmse} for NCSN++. The model’s trainable parameters amounted to 65.7~MB.
We adopted $\mu=0$, i.e., the distortionless setting, for the target generation. We did not apply the post-filtering used in the GSS approach, consisting of post-masking and Blind Analytic Normalization (BAN) \cite{BAN}, before feeding the PMWF output into the NN. Instead, we applied the post-filtering to the NN output. 

For evaluation, we used the CHiME6 dev and eval sets based on the oracle diarization labels, as well as the CHiME8 eval set based on estimated diarization labels derived for it \cite{Tawara2025}. The baseline GSS and ASR systems provided for the challenge \cite{CHiME8_2024} were used in our experiments.

Throughout training and evaluation, we applied Weighted Prediction Error (WPE) dereverberation \cite{WPE,MIMO-WPE} separately to CMs and DMs as preprocessing. Consistent with the CHiME8 baseline system, this ensured that the generated training targets were aligned with the baseline conditions.

\subsection{Compared methods}
We compared the NN for SE trained using the C2D projection (C2D) against the GSS approach (GSS) and All-to-Distant microphone projection-based SE (A2D). GSS is a strong SE front-end adopted by top-scoring systems in the challenge. Surpassing its performance can be regarded as achieving the state-of-the-art.

A2D is a variant of C2D that incorporates both CM and DM signals as inputs to the projection. Its projection matrix can be obtained by extending the CM components in Eq.~(\ref{eq:c2d}) to include all CM and DM components. Notably, A2D is equivalent to applying GSS across the full set of CM and DM signals. Compared to C2D, A2D could offer both benefits and drawbacks: while the use of all CM and DM signals enables more comprehensive modeling of spatial characteristics, the additional DM signals also introduce noise, making the projection process more challenging.

For reference, we also display the scores obtained by an NN for SE, which was trained using the CM signals directly as the training targets (CM training) without any projection. 


\subsection{Results under matched conditions with CHiME6}

\begin{table}[t]
\begin{minipage}[b]{\linewidth}
\centering
\setlength{\tabcolsep}{6.7pt}
\caption{Results under matched conditions with CHiME6 dev and eval sets using oracle diarization labels}\label{tab:dev-oracle}
\begin{tabular}{lcccccc}\toprule
 & Post filter & \multicolumn{2}{c}{tcpWER [\%] $\downarrow$} && \multicolumn{2}{c}{DNSMOS $\uparrow$} \\\cline{3-4}\cline{6-7}
 & & dev & eval && dev & eval \\\hline
 CM training & - & 56.31 & 63.15 && 1.99 & 2.12 \\
\hline
GSS & - & 20.94 & 27.37 && 2.15 & 2.07 \\
A2D & - & 20.94 & 27.11 && 2.18 & 2.10 \\
C2D (proposed)\hspace{-1cm} & - & \bf{20.13} & \bf{26.54} && \bf{2.32} & \bf{2.20} \\\hline
GSS & \checkmark & 20.01 & 26.26 && 2.25 & 2.13 \\
C2D (proposed)\hspace{-1cm} & \checkmark & \bf{19.45} & \bf{25.05} && \bf{2.39} & \bf{2.25} \\\bottomrule
\end{tabular}
\end{minipage}
%
\begin{minipage}[b]{\linewidth}
\centering
\setlength{\tabcolsep}{1.4pt}
\caption{tcpWERs $\downarrow$ 
on different scenarios under mismatched conditions with CHiME8 eval set using estimated diarization labels}\label{tab:all-scenario}
\begin{tabular}{lcccccc}\toprule
 & Post filter & chime6 & dipco & mixer6 & notsofar1 & average\\\hline
GSS & - & 38.43 & 32.86 & \bf{20.04} & 24.05 & 28.84 \\
C2D  (proposed)\hspace{-0.1cm} & - & \bf{37.57} & \bf{30.38} & 20.91 & \bf{20.70} & \bf{27.39} \\\hline
GSS & \checkmark & 37.26 & 28.22 & \bf{16.16} & 20.60 & \bf{25.56} \\
C2D (proposed)\hspace{-0.1cm} & \checkmark & \bf{36.81} & \bf{28.13} & 18.89 & \bf{20.10} & 25.98\\\bottomrule
\end{tabular}
\end{minipage}
%
\end{table}

Table~\ref{tab:dev-oracle} presents evaluation results on the CHiME6 dev and eval sets using oracle diarization labels. Performance was measured using the Time-Constrained minimum Permutation Word Error Rate (tcpWER) \cite{tcpwer} and the overall metric of the Deep Noise Suppression Mean Opinion Score (DNSMOS) \cite{reddy2021dnsmos}. The tcpWER is a multi-speaker variant of WER that accounts not only for word recognition errors but also for errors in speaker identification and utterance timing. The table also compares the performance of GSS and C2D, both with and without post-filtering, which consists of post-masking and BAN. As described in Section~\ref{sec:acond}, the post-filter was applied to the NN output for C2D.

First, the tcpWER for CM training, where the NN for SE was trained using CM signals as targets, was extremely poor. This confirms that CM signals are unsuitable as direct training targets.

In comparisons among GSS, A2D, and C2D without post-filtering, A2D provided only marginal improvement over GSS, whereas C2D achieved relative error reductions of 3.87\% and 3.13\% for the dev and eval sets, respectively. With post-filtering, both GSS and C2D further reduced tcpWER, with C2D maintaining advantages of 2.80\% and 4.61\% relative error reductions over GSS.

DNSMOS scores exhibited a similar trend, further demonstrating the effectiveness of the proposed method in improving perceptual speech quality.

Overall, these results clearly demonstrate the superiority of C2D over GSS and A2D for the CHiME6 scenario under oracle diarization when training and test conditions for the projection are well matched.

\subsection{Results under mismatched conditions with CHiME8}

Table~\ref{tab:all-scenario} presents the tcpWERs obtained across all CHiME8 scenarios using estimated diarization labels, along with their macro averages. The evaluation considers two types of mismatch between training and test for the proposed method: in diarization labels (oracle vs. estimated) and in diverse recording scenarios. Note that the NN with C2D was trained exclusively on the CHiME6 train set using oracle diarization labels. Additionally, we did not jointly trained the NN with the ASR backend unlike \cite{MITROFANOV2025101780}.

Despite the substantial mismatch between training and test conditions, C2D maintained slight superiority over GSS in most scenarios, except for Mixer6. These results demonstrate that the proposed method exhibits a certain degree of robustness against the considered mismatches. The performance degradation observed for Mixer6 may be attributed to its DM configuration being substantially different from those of the other scenarios \cite{CHiME8_2024}.

Further improving robustness to such mismatches remains an important direction for future work.


\section{Concluding remarks}
\label{sec:conclusion}
This paper introduced C2D projection, a technique for generating paired distorted and reference speech signals to train NNs for real-world SE. Given simultaneous recordings from CMs and DMs, the C2D projection optimally estimates reference signals aligned with the DM recordings.
Experiments on the CHiME6 ASR task demonstrated that an NN trained with C2D-projected data, which uses the GSS-based PMWF output as an auxiliary input, outperformed the state-of-the-art SE method for this task, namely the GSS approach itself. Moreover, in the CHiME8 ASR task, which features diverse recording scenarios, the proposed method still outperformed the GSS approach across most scenarios, despite substantial mismatches between training and test conditions.
Future work will include enhancing robustness under mismatched conditions and examining NN models that do not depend on the output of the GSS approach.

\let\oldthebibliography\thebibliography
\renewcommand{\thebibliography}[1]{%
 \oldthebibliography{#1}%
  \setlength{\itemsep}{0.0pt}%
  \setlength{\parskip}{0.0pt}%
}

\end{document}